\documentclass[showpacs,aps,onecolumn]{revtex4}
\usepackage{graphicx,graphics,color,epsfig}
\usepackage{bm}
\usepackage{amsmath}
\usepackage{amssymb}

\begin{document}

\title{Two-dimensional gapless spin liquids in frustrated SU($N$) quantum
magnets}
\author{Peng Li and Shun-Qing Shen}
\affiliation{Department of Physics, The University of Hong Kong, Pokfulam Road, Hong
Kong, China}
\date{\today}

\begin{abstract}
A class of the symmetrically frustrated SU($N$) models is constructed for
quantum magnets based on the generators of SU($N$) group. The total
Hamiltonian lacks SU($N$) symmtry. A mean field theory in the quasi-particle
representation is developed for spin liquid states. Numerical solutions in
two dimension indicate that the ground states are gapless and the
quasi-particles are Dirac particles. The mechanism may be helpful in
exploring the spin liquid phases in the spin-1 bilinear-biquadratic model
and the spin-orbital model in higher dimensions.
\end{abstract}

\maketitle

\section{Introduction}

Exploring spin liquid has been a recurrent theme in the research of strongly
correlated electron systems over the past decades \cite%
{Dagotto96,npb0,npb1,Shen98,Wen02,Sachdev03,Senthil04,Batista04}. The
picture for spin liquid is based on the concept of resonating valence bonds
(RVB), which says all spins form spin singlet pairs \cite%
{Anderson73,Anderson2,Anderson3}. Spin singlet pairs consist of either
nearest neighboring spins or long-range separated spins. Quantum coherence
of spin singlets determines the quantum properties of the state. It is
realized that a short-range RVB state usually exhibits a finite gap for spin
excitations \cite{Rokhsar88}, while a long-range RVB state possesses
antiferromagnetic long-range order (LRO) \cite{Liang88,Liang2}. Nowadays a
lot of systems are found to exhibit the behaviors of spin liquid. One route
to explore spin liquid is to focus on geometrically frustrated spin systems
or systems with breaking translational invariance. The formers include the
Kagome lattice model and the $J_{1}-J_{2}$ model. The latters include the
spin Peierls systems and the plaquette RVB systems. In these systems the
quantum and/or geometric frustration is the key point to realize spin
liquid. Another route is to deal with a series of low-dimensional spin
systems with Haldane gap or the spin systems with orbital degeneracy which
enhances quantum frustration. Such systems neither break the translational
invariance nor exhibit geometrical frustration.

In this paper we propose a class of quantum frustrated SU($N$) models for
quantum magnets, and develop a mean field theory for the spin liquid
involved in it. The models are constructed by the SU($N$) generators, but
has no SU($N$) symmetry because of the competition between the generators in
two different representations. For small $N$ it is shown that these models
are equivalent to several physical models which have been extensively
studied, and fall into the same mathematical structures in the particle
representation. To explore the spin liquid a mean field theory based on the
RVB state is developed. In two dimension ($2D$) it is found that the ground
states for $N=3$ and $4$ are gapless spin liquid and the collective
excitations are Dirac particles. As examples we apply the theory to spin 1
bilinear-biquadratic system and spin $1/2$ systems with double orbital
degeneracy.

\section{Models}

Let's start from a class of Hamiltonian
\begin{equation}
H=J_{1}\sum_{\left\langle ij\right\rangle ,\mu \nu }J_{\nu }^{\mu
}(r_{i})J_{\mu }^{\nu }(r_{j})-J_{2}\sum_{\left\langle ij\right\rangle ,\mu
\nu }J_{\nu }^{\mu }(r_{i})J_{\nu }^{\mu }(r_{j}),  \label{general}
\end{equation}%
which is defined on a lattice. $J_{\nu }^{\mu }(r_{i})$ are the $N^{2}-1$
generators of SU($N$) group, and satisfy the algebra
\begin{equation}
\left[ J_{\beta }^{\alpha }(r_{i}),J_{\nu }^{\mu }(r_{j})\right] =\delta
_{ij}\left( \delta _{\nu }^{\alpha }J_{\beta }^{\mu }(r_{i})-\delta _{\beta
}^{\mu }J_{\nu }^{\alpha }(r_{i})\right) .
\end{equation}%
We choose $J_{\nu }^{\mu }(r_{i})$ as the fundamental representation with a
single box in Young tableau. $J_{1}$ and $J_{2}$ are two coupling constants.

The first term in Eq.(\ref{general}) possesses the SU($N$) symmetry because
the operator, $\sum_{\mu \nu }J_{\nu }^{\mu }(r_{i})J_{\mu }^{\nu
}(r_{j})\equiv P_{ij}$, serves as the permutation operator, which swaps two
quantum states at sites $i$ and $j$. This term in various forms has been
studied extensively. The case for $N=2$ is the well known Heisenberg model,
which was solved exactly and exhibits a gapless ground state in one
dimension ($1D$) \cite{Sutherland75}. While in higher dimensions it is well
established that the ground state possesses antiferromagnetic LRO on
hypercubic lattices. The case for $N=3$ was known as the exchange model for
ferromagnetism \cite{Allan67,Allan2,Allan3}. For $N=4$ it was proposed that
a spin liquid exists on the triangular lattice \cite{Li98,Li2,Li3}, but a
Schwinger boson theory gives a state with LRO \cite{Shen02a}.

The second term in Eq.(\ref{general}) also has the SU($N$) symmetry on a
bipartite lattice, which means that $J_{2}$ connects two lattice sites
belonging to two different sublattices \cite%
{Arovas88,Read90,Santoro99,Zhang01,Zhang2,Harada03}. In this case the
generators at two sublattice sites are expressed in conjugate
representations since $J_{\mu }^{\nu }(r_{j})=\left[ J_{\nu }^{\mu }(r_{j})%
\right] ^{\dag }$, and thus the SU($N$) symmetry survives. In $1D$ the
ground state is gaped for $N=4$. In $2D$, the ground states are found to
exhibit the N\'{e}el-type LRO with broken SU($N$) symmetry for $N\leq 4$,
and possibly to be a spin liquid for $N>4$ \cite%
{Santoro99,Zhang01,Zhang2,Harada03}.

The mixture of the two terms makes the whole Hamiltonian in Eq. (\ref%
{general}) deviate from the two SU($N$) symmetries. We propose that this
symmetric frustration may also lead to spin liquid. To study the phase
diagram of the model in Eq.(\ref{general}) we introduce a set of creation
and annihilation operators to rewrite the Hamiltonian in the second
quantization representation. In the SU($N$) representation each site has $N$%
\ quantum states $\left\vert \mu \right\rangle $ so that we may introduce $N$
pairs of operators $b_{i\mu }^{\dag }$ and $b_{i\mu }$: $\left\vert i,\mu
\right\rangle =b_{i\mu }^{\dag }\left\vert 0\right\rangle $ with the vacuum
state $\left\vert 0\right\rangle $ at site $i$. In this way we can construct
the operator, $J_{\nu }^{\mu }(r_{i})\equiv b_{i\nu }^{\dag }b_{i\mu }$,
with a constraint for single occupancy, $\sum_{\mu =1}^{N}b_{i\mu }^{\dag
}b_{i\mu }=1$, on each site. This is the so-called hard-core condition even
if the particles are bosons. Interestingly it is found that the generators
satisfy the SU($N$) algebra in Eq. (2) for either Boson or Fermion
representation. So the Hamiltonian is reduced to
\begin{equation}
H=J_{1}\sum_{ij}P_{ij}-J_{2}\sum_{ij}B_{ij}^{\dag }B_{ij}+\sum_{i}\lambda
_{i}\left( \sum\limits_{\mu }b_{i\mu }^{\dag }b_{i\mu }-1\right) .  \label{H}
\end{equation}%
where the bond pairing operator $B_{ij}=\sum_{\mu }b_{j\mu }b_{i\mu }$ and
the Lagrangian multipliers $\lambda _{i}$ are introduced to realize the
constraint of single occupancy. The permutation operator can be expressed as
$P_{ij}=\sum_{\mu \nu }b_{i\mu }^{\dag }b_{i\nu }b_{j\nu }^{\dag }b_{j\mu
}=-\varsigma +\varsigma F_{ij}^{\dag }F_{ij}$ with $F_{ij}=\sum_{\mu
}b_{j\mu }^{\dag }b_{i\mu }$, and $\varsigma =1$ for bosons and $-1$ for
fermions. Several physical systems are shown to belong to this Hamiltonian.
It has been known for a long time that the spin SU(2) operators with $S$ can
be used to express the SU($N$) generators. Schr\"{o}dinger developed an
expression for the permutation operator in terms of $(\mathbf{S}_{i}\cdot
\mathbf{S}_{j})^{m}$ for $m=0,1,\cdots ,2S$ \cite{Schrodinger41}. (Please
see in Appendix A.) We present several examples which have the equivalent
form in either the Fermion or Boson representation.

For $N=2$ the model in Eq. (\ref{general}) is equivalent to the spin 1/2 $%
XXZ $ model,
\begin{equation}
H=\sum_{ij}\left[
J_{xx}S_{i}^{x}S_{j}^{x}+J_{xx}S_{i}^{y}S_{j}^{y}+J_{zz}S_{i}^{z}S_{j}^{z}%
\right] .
\end{equation}%
By defining $b_{1}^{\dag }\left\vert 0\right\rangle $ and $b_{2}^{\dag
}\left\vert 0\right\rangle $ as the two eigenstates of the operator $%
S_{i}^{x}$, the spin operators $S_{i}^{\alpha }$ can be expressed in terms
of $b$ operators,
\begin{subequations}
\begin{eqnarray}
S_{i}^{x} &=&\frac{1}{2}(b_{i1}^{\dag }b_{i1}-b_{i2}^{\dag }b_{i2}), \\
S_{i}^{y} &=&\frac{1}{2}(b_{i2}^{\dag }b_{i1}+b_{i1}^{\dag }b_{i2}), \\
S_{i}^{z} &=&\frac{i}{2}(b_{i2}^{\dag }b_{i1}-b_{i1}^{\dag }b_{i2}).
\end{eqnarray}
In this way, one can show that the model has the form of Eq. (\ref{H}) for $%
N=2$ with $J_{1}=(J_{zz}+J_{xx})/4$ and $J_{2}=(J_{zz}-J_{xx})/4$. The model
with a similar form was studied by Leone and Zimanyi \cite{Leone94}. When $%
J_{zz}=J_{xx}$, it returns to the well-known Heisenberg model.

For $N=3$ it is the spin 1 bilinear-biquadratic model,
\end{subequations}
\begin{equation}
H=\sum_{ij}\left[ \cos \phi \ \mathbf{S}_{i}\cdot \mathbf{S}_{j}+\sin \phi \
(\mathbf{S}_{i}\cdot \mathbf{S}_{j})^{2}\right] ,  \label{spin-1}
\end{equation}%
which is one of the prototype models exhibiting Haldane gap in $1D$
antiferromagnet \cite{Haldane83,Haldane2}. The phase diagram for $1D$ is
well established that the energy gap persists in the wide range, $-\pi
/4<\phi <\pi /4$ \cite{Affleck86,Affleck2,Affleck3}. The phase diagram for $%
2D$ and $3D$ was studied by means of quantum Monte-Carlo \cite{Harada02}.
Non-zero quadrupole moment at zero temperatures is found at the region, $%
-\pi <\phi <0$. For spin 1, each site has three states $\left\vert
m_{i}\right\rangle $ with $m_{i}=-1,0,+1$ ($i=1,2,3$) according to the
eigenvalues of $S_{i}^{z}$. We reorganize the three states and define three
operators,

\begin{subequations}
\begin{eqnarray}
b_{1}^{\dag }\left\vert 0\right\rangle &=&\frac{i}{\sqrt{2}}\left(
\left\vert m_{1}\right\rangle +\left\vert m_{3}\right\rangle \right) , \\
b_{2}^{\dag }\left\vert 0\right\rangle &=&\left\vert m_{2}\right\rangle , \\
b_{3}^{\dag }\left\vert 0\right\rangle &=&\frac{1}{\sqrt{2}}\left(
\left\vert m_{1}\right\rangle -\left\vert m_{3}\right\rangle \right) .
\end{eqnarray}%
In terms of $b$ operators, the three spin operators can be written as
\end{subequations}
\begin{subequations}
\begin{eqnarray}
S_{i}^{x} &=&i(b_{i2}^{\dag }b_{i1}-b_{i1}^{\dag }b_{i2}), \\
S_{i}^{y} &=&i(b_{i3}^{\dag }b_{i2}-b_{i2}^{\dag }b_{i3}), \\
S_{i}^{z} &=&i(b_{i1}^{\dag }b_{i3}-b_{i3}^{\dag }b_{i1}).
\end{eqnarray}%
On this basis the Hamiltonian Eq. (\ref{spin-1}) is reduced to the form of
Eq. (\ref{H}) for $N=3$ with $J_{1}=\cos \phi $ and $J_{2}=\sqrt{2}\cos
\left( \phi +\pi /4\right) $ \cite{Batista04}. At the points, $\phi =-3\pi
/4,$ $\pi /4$, $\pm \pi /2$ the model possesses the SU(3) symmetry \cite%
{Sutherland75}. It is solvable at $\tan \phi =1/3$ and the ground state is a
valence bond solid \cite{Affleck87}.

For $N=4$, we can present a model in terms of operators $S=3/2$ just like
the cases for $N=2$ and $3$. Alternatively we present the spin-orbital model
\cite{Kugel73},

\end{subequations}
\begin{equation}
H=J\sum_{ij}\left( \mathbf{S}_{i}\cdot \mathbf{S}_{j}+\mathbf{T}_{i}\cdot
\mathbf{T}_{j}\right) +4V\sum_{ij}\left( \mathbf{S}_{i}\cdot \mathbf{S}%
_{j}\right) \left( \mathbf{T}_{i}\cdot \mathbf{T}_{j}\right) ,
\label{spin-orbital}
\end{equation}%
with spin\ $S=1/2$ and orbital $T=1/2$, where both of the operators, $%
\mathbf{S}$ and $\mathbf{T}$, satisfy the SU(2) algebra. This model was used
extensively to describe the orbital physics in transition metal oxides \cite%
{Tokura00}. At $J=V$ the model has the SU(4) symmetry. There are
four simultaneous eigenstates for spin $S^{z}$ and $T^{z}$ at
each site, $\left\vert S_{i}^{z}=\pm \frac{1}{2},T_{i}^{z}=\pm \frac{1}{2}%
\right\rangle $. So we introduce four creation operators as follows,

\begin{subequations}
\begin{eqnarray}
b_{1}^{\dag }\left\vert 0\right\rangle &=&\frac{1}{\sqrt{2}}\left(
\left\vert +\frac{1}{2},-\frac{1}{2}\right\rangle -\left\vert -\frac{1}{2},+%
\frac{1}{2}\right\rangle \right) ; \\
b_{2}^{\dag }\left\vert 0\right\rangle &=&\frac{i}{\sqrt{2}}\left(
\left\vert +\frac{1}{2},-\frac{1}{2}\right\rangle +\left\vert -\frac{1}{2},+%
\frac{1}{2}\right\rangle \right) ; \\
b_{3}^{\dag }\left\vert 0\right\rangle &=&\frac{i}{\sqrt{2}}\left(
\left\vert +\frac{1}{2},+\frac{1}{2}\right\rangle -\left\vert -\frac{1}{2},-%
\frac{1}{2}\right\rangle \right) ; \\
b_{4}^{\dag }\left\vert 0\right\rangle &=&\frac{1}{\sqrt{2}}\left(
\left\vert +\frac{1}{2},+\frac{1}{2}\right\rangle +\left\vert -\frac{1}{2},-%
\frac{1}{2}\right\rangle \right) .
\end{eqnarray}%
By introducing a four component spinor $\Psi ^{\dag }=\left( b_{1}^{\dag
},b_{2}^{\dag },b_{3}^{\dag },b_{4}^{\dag }\right) $, we can express the
spin and orbital operators in terms of $\Psi ^{\dag }$ and $\Psi $,
\end{subequations}
\begin{subequations}
\begin{eqnarray}
S^{x} &=&\frac{1}{2}\Psi ^{\dag }\left( \sigma _{y}\otimes \sigma
_{0}\right) \Psi ,\text{ } \\
S^{y} &=&\frac{1}{2}\Psi ^{\dag }\left( \sigma _{x}\otimes \sigma
_{y}\right) \Psi ,\text{ } \\
S^{z} &=&\frac{1}{2}\Psi ^{\dag }\left( -\sigma _{z}\otimes \sigma
_{y}\right) \Psi ,\text{ } \\
T^{x} &=&\frac{1}{2}\Psi ^{\dag }\left( -\sigma _{y}\otimes \sigma
_{z}\right) \Psi , \\
T^{y} &=&\frac{1}{2}\Psi ^{\dag }\left( -\sigma _{y}\otimes \sigma
_{x}\right) \Psi , \\
T^{z} &=&\frac{1}{2}\Psi ^{\dag }\left( \sigma _{0}\otimes \sigma
_{y}\right) \Psi .
\end{eqnarray}%
Once again one can show that the spin-orbital model falls into the general
form of Hamiltonian as Eq. (\ref{H}) with $J_{1}=(J+V)/2$ and $J_{2}=(J-V)/2$%
.

So far we have shown that the three physical models can be expressed in a
general form in terms of SU($N$) generators as Eq.(1) or in the particle
representation as Eq. (2). For larger $N$, the models are related to some
high spin systems with orbital degeneracy, for example, the spin-1 system
with double orbital degeneracy \cite{Shen02-prl}.

\section{Mean Field Theory}

Now we turn to the phase diagram of the model on a $d$-dimensional
hypercubic lattice. We may choose the operators, $b$'s, as either bosons, $%
[b_{i\mu },b_{j\nu }^{\dag }]=\delta _{ij}\delta _{\mu \nu }$, or fermions, $%
\left\{ b_{i\mu },b_{j\nu }^{\dag }\right\} =\delta _{ij}\delta _{\mu \nu }$%
. In principle the bosons tend to condensate to the lowest energy state at
low temperatures and to form a quantum ordered state, while the fermions
tend to form a Fermi sea and a quantum disordered states. In practice it is
inevitable to introduce adequate approximation schemes to deal with a
many-body system in either the Bose or Fermion representation. Since our
purpose is to search the regime of the spin liquid phase, we choose the
Fermion representation in this paper. Without loss of generality we can
assume that the system is defined on a $d$-dimensional simple cubic lattice,
and take $J_{1},J_{2}>0$. In this way the Hamiltonian is semi-negative in
the Fermi representation. Now the second term in Eq.(\ref{general}) plays a
role of an attractive interaction between pairing valence bonds.\cite%
{Zhang01} The Hubbard-Stratonovich transformation is performed to decouple
the Hamiltonian into a bilinear form. Two types of mean fields are
introduced,
\end{subequations}
\begin{subequations}
\begin{eqnarray}
\Delta _{1}(k) &=&J_{1}\sum_{\delta }\left\langle F_{i,i+\delta
}\right\rangle e^{ik\cdot \delta }\equiv 2J_{1}F\sum_{\alpha }\cos k_{\alpha
}; \\
\Delta _{2}(k) &=&J_{2}\sum_{\delta }\left\langle B_{i,i+\delta
}\right\rangle e^{ik\cdot \delta }\equiv 2J_{2}B\sum_{\alpha }\sin k_{\alpha
}
\end{eqnarray}%
where $F=\left\langle F_{i,i+\delta }\right\rangle =\left\langle
F_{i,i-\delta }\right\rangle $ and $B=i\left\langle B_{i,i+\delta
}\right\rangle =-i\left\langle B_{i,i-\delta }\right\rangle ,$ and $\alpha
=x,y,\cdots .$ The bracket $\left\langle \cdots \right\rangle $ represents
the thermodynamic average. The chemical potential $\lambda _{i}$ is taken to
be site-independent, $\lambda _{i}=\lambda ,$ which can be also regarded as
a mean field. In the momentum space the mean field Hamiltonian is
\end{subequations}
\begin{eqnarray}
H &=&\sum_{k,\mu }\epsilon (k)b_{k\mu }^{\dag }b_{k\mu }-\frac{1}{2}%
\sum_{k}\Delta _{2}(k)\left( b_{k\mu }b_{-k\mu }+b_{-k\mu }^{\dag }b_{k\mu
}^{\dag }\right)  \notag \\
&&-\lambda N_{\Lambda }+dN_{\Lambda }J_{1}F^{2}+dN_{\Lambda }J_{2}B^{2}.
\end{eqnarray}%
where $\epsilon (k)=\lambda -\Delta _{1}(k)$, $N_{\Lambda }$ is the total
number of lattice sites. This is a Bardeen-Cooper-Schrieffer (BCS) type of
Hamiltonian for $N$-component degenerate fermions. By performing the
Bogoliubov transformation,
\begin{equation}
\gamma _{k\mu }=u_{k}b_{k\mu }-v_{k}b_{-k\mu }^{\dag };\text{ }\gamma
_{-k\mu }^{\dag }=u_{k}b_{-k\mu }^{\dag }+v_{k}b_{k\mu }
\end{equation}%
with the coherence factors satisfying

\begin{subequations}
\begin{eqnarray}
u_{k}^{2} &=&\frac{1}{2}\left[ 1+\frac{\epsilon (k)}{\omega (k)}\right] , \\
v_{k}^{2} &=&\frac{1}{2}\left[ 1-\frac{\epsilon (k)}{\omega (k)}\right] , \\
2u_{k}v_{k} &=&\frac{\Delta _{2}(k)}{\omega (k)},
\end{eqnarray}%
one can diagonalize the Hamiltonian as
\end{subequations}
\begin{equation}
H=\sum_{k,\mu }\omega (k)\gamma _{k\mu }^{\dag }\gamma _{k\mu }+E_{0},
\end{equation}
where the spectrum and the ground energy are
\begin{equation}
\omega (k)=\sqrt{\epsilon (k)^{2}+\Delta _{2}^{2}(k)},
\end{equation}
\begin{equation}
E_{0}=-\frac{N}{2}\sum_{k}\omega (k)+\frac{N-2}{2}\lambda N_{\Lambda
}+dN_{\Lambda }J_{1}F^{2}+dN_{\Lambda }J_{2}B^{2}.
\end{equation}
The spectra of $\omega (k)$ are $N$-fold degenerate for the quasiparticles.
From the free energy for the fermion gas
\begin{equation}
\Omega =-\frac{N}{\beta }\sum_{k}\ln (1+e^{-\beta \omega (k)})+E_{0},
\end{equation}
we obtain a set of the mean field equations by optimizing the free energy
with respect to the mean fields $F$, $B$, and $\lambda$,
\begin{subequations}
\begin{eqnarray}
\int \frac{dk}{\left( 2\pi \right) ^{d}}\frac{\epsilon (k)}{\omega (k)}\tanh
\frac{\beta \omega (k)}{2} &=&\frac{N-2}{N}; \\
\int \frac{dk}{\left( 2\pi \right) ^{d}}\frac{\epsilon (k)\left(
-\sum_{\alpha }\cos k_{\alpha }\right) }{\omega (k)}\tanh \frac{\beta \omega
(k)}{2} &=&\frac{2dF}{N}; \\
\int \frac{dk}{\left( 2\pi \right) ^{d}}\frac{\Delta _{2}(k)\left(
\sum_{\alpha }\sin k_{\alpha }\right) }{\omega (k)}\tanh \frac{\beta \omega
(k)}{2} &=&\frac{2dB}{N}.
\end{eqnarray}%
Thus the mean field Hamiltonian is solved together with the self-consistent
equations for the three types of mean fields.

We have $N$ degenerate spectra for quasi-fermions. These are not the
collective modes, and cannot be measured explicitly. In order to calculate
spin susceptibility we define the Matsubara Green's function in the form of
a $2\times 2$ matrix,
\end{subequations}
\begin{equation}
G_{\mu }(k,\tau )=-\left\langle T_{\tau }\left(
\begin{array}{cc}
b_{k\mu }(\tau )b_{k\mu }^{\dag }(0) & b_{k\mu }(\tau )b_{-k\mu }(0) \\
b_{-k\mu }^{\dag }(\tau )b_{k\mu }^{\dag }(0) & b_{-k\mu }^{\dag }(\tau
)b_{-k\mu }(0),%
\end{array}%
\right) \right\rangle
\end{equation}%
where the bracket $\left\langle \cdots \right\rangle $ means that the
thermodynamic average is made, and the factor $T_{\tau }$ is the imaginary
time order operator. The Green's function of the frequency is obtained as
follows,
\begin{equation}
G_{\mu }(k,i\omega _{n})=\frac{i\omega _{n}\sigma _{0}+\epsilon (k)\sigma
_{z}+\Delta _{2}(k)\sigma _{x}}{\left( i\omega _{n}\right) ^{2}-\omega
^{2}(k)},
\end{equation}%
where $\omega _{n}=(2n+1)\pi /\beta $ for all integer $n$. Due to the
degeneracy of the spectra the Green's functions are independent of the index
$\mu$. The imaginary time dynamic correlation function for $J_{\nu }^{\mu }$
is defined as
\begin{equation}
\chi _{\mu \nu }(q,i\omega _{n})=\int_{0}^{\beta }d\tau e^{i\omega _{n}\tau
}\left\langle T_{\tau }J_{\mu }^{\nu }(q,\tau )J_{\nu }^{\mu
}(-q,0)\right\rangle .
\end{equation}%
The imaginary part of $\chi _{\mu \nu }(q,\omega )$ is worked out as
\begin{eqnarray}
Im\chi _{\mu \nu }(q,\omega ) &=&\pi \int \frac{dk}{\left( 2\pi \right) ^{d}}%
\delta (\omega -\omega (k)-\omega (k+q))  \notag \\
&&\times \left( u_{k}^{2}v_{k+q}^{2}+u_{k}v_{k}u_{k+q}v_{k+q}\delta _{\mu
\nu }\right)
\end{eqnarray}%
for $\omega >0$ at $T=0$. It is shown that $Im\chi_{\mu \nu }(q,\omega )$
becomes non-zero only when $\omega >\Delta _{gap}=2\min \omega (k).$ Thus if
the spectra for quasiparticles have an energy gap the collective excitation
for the dynamic correlation function also have a finite energy gap. The
energy gap $\Delta _{gap}$ can be evaluated by solving the mean field
equations.

\begin{figure}[tbp]
\centerline{\psfig{figure=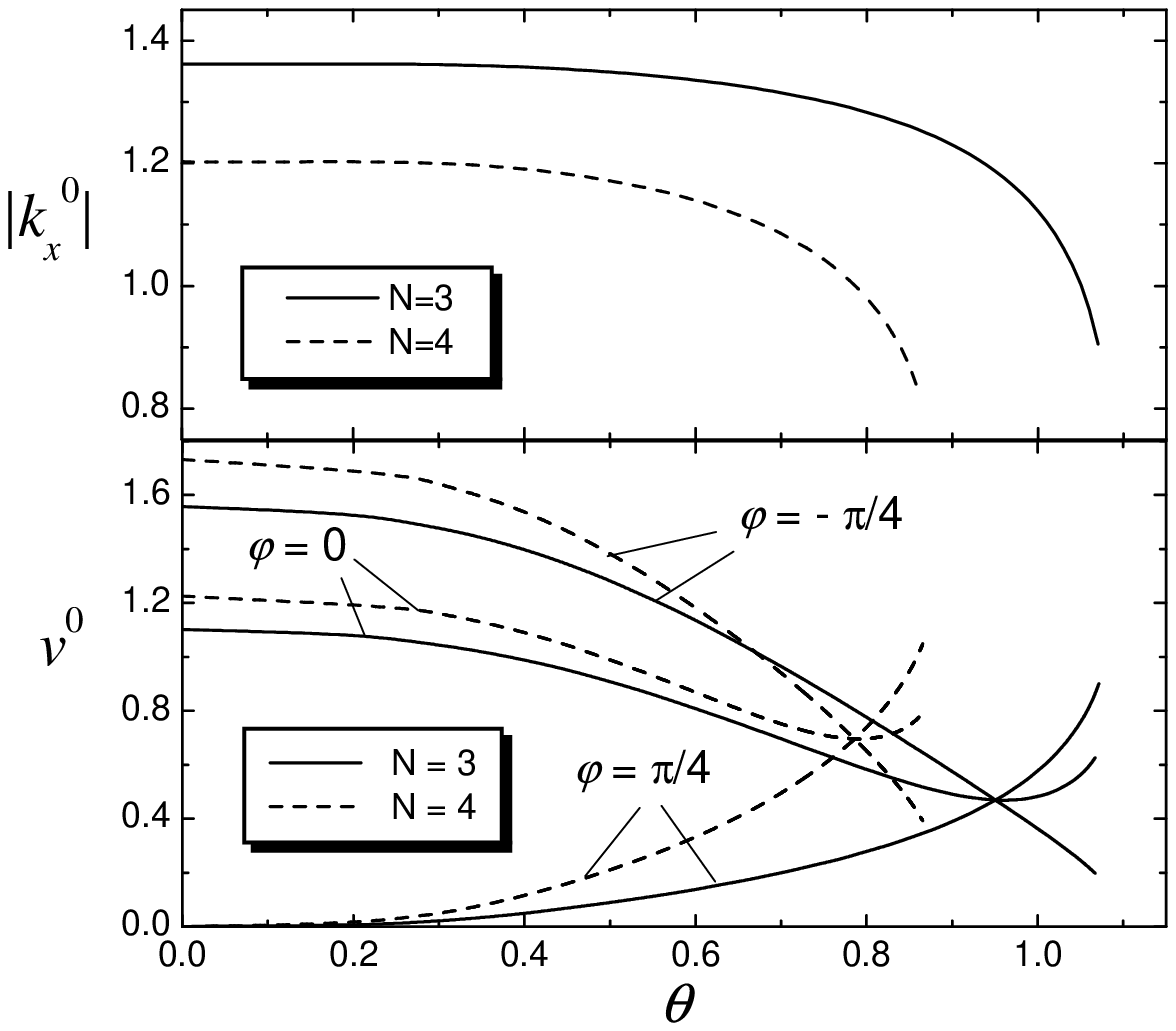,height=7cm,width=9cm}}
\caption{Above: the value of $k_{x}^{0}$ at two points for the fermi surface
($k_{x}^{0},-k_{x}^{0}$) and ($-k_{x}^{0},k_{x}^{0}$)\ for $N=3,4$ in square
lattice. Bottom: The velocities of the Dirac quasiparticles along the
different directions. For each $N$ there exists a fixed point at which the
velocity becomes isotropic. The angle $\protect\theta =\arctan (J_{2}/J_{1}).
$}
\label{FIG1}
\end{figure}

\section{Two-dimensional Gapless Spin Liquids}

It is easy to see that when $J_{2}=0$ or $B=0$ the spectra for fermions are
reduced to an ideal fermion gas. In the case the system is a type of gapless
spin liquid as discussed in the SU(2) and SU(4) systems. When $J_{2}$
increases a non-zero solution for $B$ and $F$ is obtained. We take $%
J_{1}=J\cos \theta $ and $J_{2}=J\sin \theta $ with $0<\theta <\pi /2.$ For
finite $N,$ the mean field equations are solved numerically at $T=0$. In one
dimensional case a finite gap is found in a large regime of the parameter
space. In this paper we focus on $2D$ systems. Numerical solution for
non-zero $B$ and $F$ has lower energy than other mean field solutions with $%
F=0$ or $B=0$ when $\theta <1.075$ for $N=3$ and $<0.867$ for $N=4$.
Opposite to the $1D$ cases non-zero $B$ and $F$ do not produce an gaped
phase. Instead they lead to a gapless spin liquid with $\Delta _{gap}=0$.
Notice that along the line $k_{x}^{0}=-k_{y}^{0}$, we have $\Delta _{2}(k)=0$%
. And on this line if $\epsilon (k)=0$ the spectrum becomes $\omega (k)=0$.
So we get the solution,
\begin{equation}
k_{x}^{0}=-k_{y}^{0}=\pm \arccos \frac{\lambda }{4J_{1}F}.
\end{equation}%
Numerical values of the parameters show that $\frac{\lambda }{4J_{1}F}\leq 1$
always holds for the mean field solution. $\left\vert k_{x}^{0}\right\vert $
is plotted in Fig. 1. In other words, the Fermi surface actually consists of
two points located at $k_{x}^{0}=-k_{y}^{0}=\pm \arccos \frac{\lambda }{%
4J_{1}F}$. We expand the dispersion relation near the Fermi point $\left(
k_{x}^{0},k_{y}^{0}\right) $ with
\begin{subequations}
\begin{eqnarray}
\Delta k_{x} &=&k_{x}-k_{x}^{0}=\Delta k\cos \varphi ; \\
\Delta k_{y} &=&k_{y}-k_{y}^{0}=\Delta k\sin \varphi ,
\end{eqnarray}%
the spectrum becomes linear with respect to $\Delta k,$
\end{subequations}
\begin{equation}
\omega \left( \mathbf{k}^{0}+\Delta \mathbf{k}\right) \simeq v^{0}\Delta k
\end{equation}%
where the anisotropic velocity is
\begin{equation}
v^{0}=\lambda \sqrt{2}\sqrt{c_{1}^{2}+\left( c_{2}^{2}-c_{1}^{2}\right) \sin
^{2}\left( \varphi +\pi /4\right) }
\end{equation}%
with $c_{1}^{2}=\left( 2J_{1}F/\lambda \right) ^{2}-1/4$ and$%
c_{2}^{2}=B^{2}\tan ^{2}\theta/F^{2}.$ The dispersion relation is
linear and the quasiparticles are type of Dirac particles. Its
velocity is in general anisotropic and has its minimal for
$\varphi =\pi /4$ and its maximal for $-\pi /4$ for a specific
$\theta $. At $\theta =0.951$ for $N=3$ and $0.786$ for $N=4$ the
velocity becomes \textit{isotropic}. Beyond the regime in Fig. 1
we do not anticipate the present mean field theory is still valid
because it is already known that the long-range order exists in
the ground state in $2D$ when $J_{1}=0$ \cite{Zhang01,Harada03}.

\begin{figure}[tbp]
\centerline{\psfig{figure=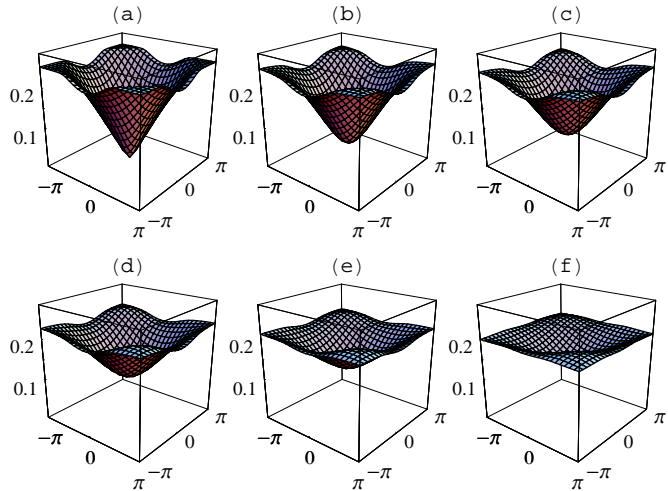,height=6.5cm,width=8.8cm}}
\caption{Diagonal susceptibility $\protect\chi _{\protect\alpha \protect%
\alpha }(q_{x},q_{y})$ for $N=4$ and different couplings $\protect\theta %
=0.300$ (a), 0.500 (b), 0.600 (c), 0.700 (d), 0.786 (e), 0.850 (f). x-axis: $%
q_{x}\in (-\protect\pi ,\protect\pi ),$ y-axis: $q_{y}\in (-\protect\pi ,%
\protect\pi )$, z-axis: $\protect\chi _{\protect\alpha \protect\alpha }$ $%
\in (0.00,$ $0.28)$.}
\label{FIG2}
\end{figure}

\begin{figure}[tbp]
\centerline{\psfig{figure=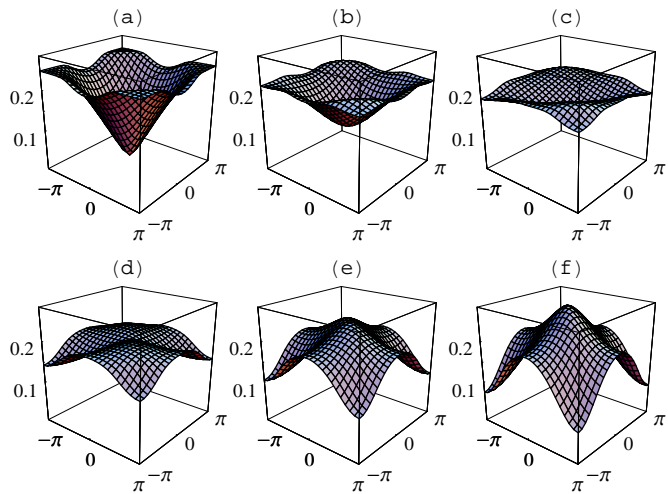,height=6.5cm,width=8.8cm}}
\caption{Off-diagonal susceptibility $\protect\chi _{\protect\alpha \protect%
\beta }$ for $N=4$ and different couplings $\protect\theta =0.300$ (a),
0.500 (b), 0.600 (c), 0.700 (d), 0.786 (e), 0.850 (f). The axis labels are
the same as in Fig. 2..}
\label{FIG3}
\end{figure}

We have also calculated the static correlation function $\chi _{\mu \mu }$
and $\chi _{\mu \nu }$ ($\mu \neq \nu $) for $N=3$ and $4$. We found that
the results for the two systems are very similar. So we only present the
results for $N=4$ here. In Fig. 2, we plot the diagonal susceptibility, $%
\chi _{\mu \mu }(q_{x},q_{y})$, for different couplings. All values of $\chi
$ are convergent, which indicates the absence of long-range order as
expected in a spin liquid state. We observe that $\chi $ has its maximal
value at points ($\pm \pi ,\pm \pi $) for small $\theta $ (or $J_{1}$),
which shows the static antiferromagnetic correlation dominates. The maximal
values decrease with the increasing $\theta $ (from Fig. 2(a) to 2(b)), and $%
\chi _{\mu \mu }$ becomes almost flat for large $\theta $. In Fig.3, we plot
the off-diagonal $\chi _{\mu \nu }$ ($\mu \neq \nu $). We found that the
maximal value of $\chi $ changes from the points ($\pm \pi ,\pm \pi $) to
(0,0) with $\theta $ increasing. Thus the antiferromagnetic correlation
decreases very quickly, and the \ ferromagnetic correlation dominates at
large $\theta $. For an intermediate $\theta $ either the diagonal or the
off-diagonal $\chi $ has a relatively flat structure, which implies that the
state is a very well-defined spin liquid.

Though the mean field theory has lots of disadvantages, it is still used
extensively to study the many body system with strong correlation after some
adequate transformations. Until present there has no definite conclusion on
whether the $2D$ SU($N$) ($N>2$) antiferromagnetic model with $J_{1}>0$ and $%
J_{2}=0$ possesses LRO ground state \cite{Li98,Li2,Li3,Shen02a}. In the
present mean field theory it is assumed that the ground state is a spin
liquid in the case. Quantum Monte Carlo simulation may provide helpful
information. The phase diagram of the $2D$ spin 1 bilinear-biquadratic model
in Eq. (\ref{spin-1}) was studied numerically for $-\pi <\phi <0$ \cite%
{Harada02}. It is found that the ground state is antiferromagnetic for $-\pi
/2<\phi <0$, which corresponds to the regime $\pi /4<\theta <\pi /2$ for $%
N=3 $ in Fig. 1. As there lacks data in the regime $0<\theta <\pi /4$ ($%
\approx 0.785$) ($0<\phi <\pi /4$) the numerical calculation does not
exclude possible existence of spin liquid in the case of $J_{1},J_{2}>0$
though the antiferromagnetic fluctuations may be very strong.

\section{Conclusion}

In conclusion, we have proposed a class of frustrated SU($N$) models for
quantum magnets based on the generators of SU($N$) group, and showed that
several physical models have the same mathematical structure in the Fermion
or Boson representation. The model we constructed breaks SU($N$) symmetry
when $J_{1}$ and $J_{2}$ are non-zero simultaneously, although the two parts
in the model do have different SU($N$) symmetries, respectively. Usually
geometric quantum frustration on lattices or coupling structures is one of
the main driving forces to generate spin liquid. The mixture of two parts
with different SU($N$) symmetries provides us a possibly new way to seek for
spin liquid in higher dimensions. In this paper a mean field theory with
paired valence bonds for quasi-particles is developed to explore this kind
of spin liquid. A gapless spin liquid exists in a large extent regime in two
dimension. It is believed that the mechanism of the spin liquid is the
competition of two interactions with different SU($N$) symmetries. Thus the
symmetric frustration is a new possible route to explore the spin liquid in
higher dimensions.

\begin{acknowledgments}
The authors thank Professor F. C. Zhang for helpful discussions. This work
was supported by the Research Grant Council of Hong Kong (Projects No. HKU
7109/02P and HKU 7038/04P).
\end{acknowledgments}

\section{Appendix:}

\subsection{Spin exchange operator}

The role of the permutation operator is to swap the two quantum states at
two different sites,
\begin{equation*}
P_{ij}\left\vert i,\alpha ;j,\beta \right\rangle =\left\vert
i,\beta ;j,\alpha \right\rangle .
\end{equation*}%
$P_{ij}^{2}=1$ and its two eigenvalues are $\pm 1.$ The connection
between the permutation operator and spin operator was first
established for spin 1/2 by Dirac\cite{Dirac58}, and then
generalized to the case of arbitrary spin $S$ by Schr\"{o}dinger
\cite{Schrodinger41}, who showed that
\begin{equation*}
P_{ij}=\sum_{n=0}^{2S}A_{n}\left( \mathbf{S}_{i}\cdot \mathbf{S}%
_{j}\right) ^{n}
\end{equation*}%
where the coefficients $A_{n}$ typically have the values, for $S=1/2$, $%
A_{0}=1/2, $ $A_{1}=2;$ for $S=1$, $A_{0}=-1,$ $A_{1}=1,$
$A_{2}=1;$ for $S=3/2$, $A_{0}=-67/32,$ $A_{1}=-9/8,$
$A_{2}=11/18,$ $A_{3}=2/9.$ The general expression for $P_{ij}$ is
\begin{eqnarray*}
P_{ij} &=&\left[ \left( 2S\right) !\right] ^{-2}\left[ \mathbf{M}%
-\left( 2S-1\right) 2S\right] \\
&&\times \left[ \mathbf{M}-\left( 2S-2\right) \left( 2S-1\right) \right]
\cdots \left[ \mathbf{M}-2\right] M \\
&&-\left[ \left( 2S-1\right) !\right] ^{-2}\left[ \mathbf{M}-\left(
2S-2\right) \left( 2S-1\right) \right] \\
&&\cdots \times \left[ \mathbf{M}-2\right] \mathbf{M} \\
&&+\cdots +(-1)^{2S-1}\mathbf{M}+(-1)^{2S} \\
&=&\sum_{n=0}^{2S}(-1)^{n}\left[ \left( 2S-n\right) !\right] ^{-2}\left[
\mathbf{M}-\left( 2S-1-n\right) \left( 2S-n\right) \right] \\
&&\cdots \times \left[ \mathbf{M}-2\right] \mathbf{M}
\end{eqnarray*}%
where
\begin{equation*}
\mathbf{M}=\left( \mathbf{S}_{i}+\mathbf{S}_{j}\right) ^{2}.
\end{equation*}

\subsection{Projection operators for two spins}

Using $\mathbf{S}_{ij}\equiv \mathbf{S}_{i}\cdot \mathbf{S}_{j}=\frac{1}{2}%
\left( \mathbf{S}_{i}+\mathbf{S}_{j}\right) ^{2}-S(S+1)$ and the
completeness relation for the projection operators $%
\sum_{J=0}^{2S}P_{J}^{S}(i,j)=1$, one finds
\begin{equation*}
\left( \mathbf{S}_{ij}\right) ^{m}=\sum_{J=0}^{2S}\left[ \frac{1}{2}%
J(J+1)-S(S+1)\right] ^{m}P_{J}^{S}(i,j).
\end{equation*}
where $P_{J}^{S}(i,j)$ is defined as the projection operator for total spin $%
J$\ of two spins $S$. For spin 1/2,

\begin{eqnarray*}
P_{J=0}^{S=1/2}(i,j) &=&\frac{1}{4}-\mathbf{S}_{ij}; \\
P_{J=1}^{S=1/2}(i,j) &=&\frac{3}{4}+\mathbf{S}_{ij}.
\end{eqnarray*}%
For spin 1,

\begin{eqnarray*}
P_{J=0}^{S=1}(i,j) &=&-\frac{1}{3}+\frac{1}{3}\mathbf{S}_{ij}^{2}; \\
P_{J=1}^{S=1}(i,j) &=&1-\frac{1}{2}\mathbf{S}_{ij}-\frac{1}{2}\mathbf{S}%
_{ij}^{2}; \\
P_{J=2}^{S=1}(i,j) &=&\frac{1}{3}+\frac{1}{2}\mathbf{S}_{ij}+\frac{1}{6}%
\mathbf{S}_{ij}^{2}.
\end{eqnarray*}%
For spin 3/2,
\begin{eqnarray*}
P_{J=0}^{S=3/2}(i,j) &=&\frac{33}{128}+\frac{31}{96}\mathbf{S}_{ij}\mathbf{-}%
\frac{5}{72}\mathbf{S}_{ij}^{2}\mathbf{-}\frac{1}{18}\mathbf{S}_{ij}^{3}; \\
P_{J=1}^{S=3/2}(i,j) &=&-\frac{81}{128}-\frac{117}{160}\mathbf{S}_{ij}%
\mathbf{+}\frac{9}{40}\mathbf{S}_{ij}^{2}\mathbf{+}\frac{1}{10}\mathbf{S}%
_{ij}^{3}; \\
P_{J=2}^{S=3/2}(i,j) &=&\frac{165}{128}+\frac{23}{96}\mathbf{S}_{ij}\mathbf{-%
}\frac{17}{72}\mathbf{S}_{ij}^{2}\mathbf{-}\frac{1}{18}\mathbf{S}_{ij}^{3};
\\
P_{J=3}^{S=3/2}(i,j) &=&\frac{11}{128}+\frac{27}{160}\mathbf{S}_{ij}\mathbf{+%
}\frac{29}{360}\mathbf{S}_{ij}^{2}\mathbf{+}\frac{1}{90}\mathbf{S}_{ij}^{3}.
\end{eqnarray*}

\end{document}